\documentclass{svproc}
\usepackage{url}

\usepackage{amssymb}
\usepackage{epsfig}
\usepackage{amsmath}
\usepackage[mathscr]{eucal}
\usepackage{subfigure}
\usepackage{graphicx}

\newcommand{\ind}{1\hspace{-2.3mm}{1}}
\renewcommand{\qed}{\hfill{\ \ \rule{2mm}{2mm}} \vspace{0.2in}}

\begin{document}
\mainmatter              % start of a contribution
\title{Approximate MDS Property of Linear Codes}
\titlerunning{Approximate MDS Property of Linear Codes}  % abbreviated title (for running head)
\author{Ghurumuruhan Ganesan\inst{1}}
% \and Roger Temam\inst{2} Jeffrey Dean \and David Grove \and Craig Chambers \and Kim~B.~Bruce \and Elsa Bertino}
%
\authorrunning{G. Ganesan} % abbreviated author list (for running head)
\tocauthor{G. Ganesan}
\institute{Institute of Mathematical Sciences, HBNI, Chennai\\
\email{gganesan82@gmail.com}}
%\and
%Universit\'{e} de Paris-Sud, Laboratoire d'Analyse Num\'{e}rique, B\^{a}timent 425,\\ F-91405 Orsay Cedex, France}

\maketitle              % typeset the title of the contribution

\begin{abstract}
In this paper, we study the weight spectrum of linear codes with \emph{super-linear} field size and use the probabilistic method to show that for nearly all such codes, the corresponding weight spectrum is very close to that of a maximum distance separable (MDS) code.

% We would like to encourage you to list your keywords within
% the abstract section using the \keywords{...} command.
\end{abstract}
\renewcommand{\theequation}{\arabic{section}.\arabic{equation}}
\setcounter{equation}{0}
\section{Introduction} \label{intro}
MDS codes have the largest possible minimum distance since they meet the Singleton bound with equality (Huffman and Pless~\cite{huff}) and many properties of the weight spectrum of MDS codes are known. For example, the weight spectrum of an MDS code is unique (Tolhuizen~\cite{tol}, MacWilliams and Sloane~\cite{macw}) and any MDS code with length~\(n\) and dimension~\(k\) has precisely~\(k\) distinct non-zero weights~\(n,n-1,\ldots,n-k+1\) (Ezerman et al.~\cite{ezer}). In this paper, we study the weight spectrum of linear codes that are not necessarily MDS but are equipped with a field size that grows super-linear in the code length. We use the probabilistic method and weight concentration properties to show that such codes closely resemble MDS codes in terms of the weight spectrum. The paper is organized as follows: In the next Section~\ref{approx_mds}, we state and prove our main result regarding the approximate MDS property of linear codes with super-linear field size.

\renewcommand{\theequation}{\arabic{section}.\arabic{equation}}
\setcounter{equation}{0}
\section{Approximate MDS property of linear codes}\label{approx_mds}
Let~\(q\) be a power of a prime number and let~\(\mathbb{F}_q\) be the finite field containing~\(q\) elements.
For integers~\(n \geq k \geq 1,\) a subset~\({\mathcal C} \subset \mathbb{F}^{n}_q\) of cardinality~\(q^{k}\) is defined to be an~\((n,k)_q-\)code.
A vector subspace of~\(\mathbb{F}^{n}_q\) of dimension~\(k\) is defined to be a \emph{linear} code and is also said to be an~\([n,k]_q-\)code.
Elements of~\({\mathcal C}\) are called codewords or simply words.

For two words~\(\mathbf{c} = (c_1,\ldots,c_n)\) and~\(\mathbf{d} = (d_1,\ldots,d_n)\) in~\(\mathbb{F}^{n}_q,\) we define the \emph{Hamming distance} between~\(\mathbf{c}\) and~\(\mathbf{d}\) to be~\(d_H(\mathbf{c},\mathbf{d}) = \sum_{i=1}^{n} \ind(c_i \neq d_i),\) where~\(\ind(.)\) refers to the indicator function. The Hamming weight of~\(\mathbf{c}\) is the number of non-zero entries in~\(\mathbf{c}.\) All distances and weights in this paper are Hamming and so we suppress the term Hamming throughout. We define the minimum distance~\(d_{H}({\mathcal C})\) of the code~\({\mathcal C}\) to be the minimum distance between any two codewords of~\({\mathcal C}.\)

From the Singleton bound, we know that~\(d_H({\mathcal C}) \leq n-k+1\) and if~\(q \geq n-1\) is a power of prime, there are~\([n,k]_q-\)codes that achieve the Singleton bound. Such codes are called maximum distance separable (MDS) codes (pp.~\(71,\) (Huffman and Pless~\cite{huff}) and the MDS conjecture asserts that~\(q=n-1\) is essentially the minimum required field size to construct MDS codes (see for example~ (Alderson~\cite{alder}), for a precise formulation).

In our main result of this paper, we show that nearly all linear codes with super-linear field size behave approximately like MDS codes. We begin with a couple of definitions. Let~\({\mathcal C}\) be a linear~\([n,k]_q-\)code and suppose for~\(1 \leq w \leq n,\) the code~\({\mathcal C}\) contains~\(A_w\) codewords of weight~\(w.\) We define the~\(n-\)tuple~\((A_1,\ldots,A_n)\) to be the \emph{weight spectrum} of~\({\mathcal C}.\) The weight spectrum of an~\([n,k]_q-\)MDS code~\({\mathcal C}\)  with~\(n \leq q\) is as follows (Theorem~\(6,\) pp. 320--321, MacWilliams and Sloane~\cite{macw}):\\
\((p1)\) For~\(1 \leq w \leq n-k,\) the number of codewords of weight~\(w\) is~\(\lambda_w := 0.\)\\
\((p2)\) For each~\(D:=n-k+1 \leq w \leq n,\) the number of codewords of weight~\(w\) equals
\begin{equation}\label{mu_w_mds}
\lambda_w := {n \choose w} (q-1) \sum_{j=0}^{w-D} (-1)^{j} {w-1 \choose j} q^{w-D-j}.
\end{equation}
It is well-known (Ezerman et al.~\cite{ezer}) that if~\(n \leq q\) then~\(\lambda_w >0\) for each~\(n-k+1 \leq w \leq n.\)

The following result shows that nearly all linear codes with super-linear field size have a weight spectra closely resembling that of an MDS code.
\begin{theorem}\label{thm_code_wt} For integer~\(n \geq 4\) let~\(k = k(n)\) be an integer and~\(q = q(n)\) be a power of a prime number satisfying
\begin{equation}\label{k_q_cond}
\frac{1}{\sqrt{\log{n}}} \leq \frac{k(n)}{n} \leq 1-\frac{1}{\sqrt{\log{n}}} \text{ and }\frac{q(n)}{n} \longrightarrow \infty
\end{equation}
as~\(n \rightarrow \infty.\) Let~\({\mathcal P}\) be the set of all~\([n,k]_q-\)codes and let~\({\mathcal Q} \subseteq {\mathcal P}\) be the set of all codes satisfying the following properties:\\
\((a1)\) There exists no word of weight~\(w\) for any~\(1 \leq w  \leq n-k-\frac{2n}{\log{q}}.\)\\
\((a2)\) For each~\(n-k+5 \leq w \leq n,\) the number of codewords of weight~\(w\) equals lies between~\(\lambda_w\left(1-\frac{3n}{q}\right)\) and~\(\lambda_w\left(1+\frac{3n}{q}\right).\)\\
For all~\(n\) large we have that
\begin{equation}\label{q_est}
\#{\mathcal Q} \geq \#{\mathcal P} \cdot \left(1-\frac{18q}{n^2}\right).
\end{equation}
%for all~\(n\) large.
\end{theorem}
From~(\ref{k_q_cond}) we have that~\(\frac{n}{\log{q}} = o(k), \frac{n}{q} = o(1)\) and so in addition if we have that~\(\frac{q}{n^2} = o(1),\) then comparing with~\((a1)-(a2),\) we see that nearly all linear codes behave approximately like an MDS code. %in the sense of~\((a1)-(a2).\)

In the following subsection, we derive a couple of preliminary estimates used in the proof of Theorem~\ref{thm_code_wt} and in the next subsection, we prove Theorem~\ref{thm_code_wt}.

\subsection*{Preliminary Estimates}
We use the probabilistic method to prove Theorem~\ref{thm_code_wt}. Let~\(\mathbf{G}\) be a random\\\(k \times n\) matrix with entries i.i.d. uniform in~\(\mathbb{F}_q.\) We prove Theorem~\ref{thm_code_wt} by estimating the weights of the words generated by the code~\({\mathcal G} := \{\mathbf{x} \cdot \mathbf{G}\}_{\mathbf{x} \in \mathbb{F}_q^{k}}.\) All vectors throughout are row vectors.

We collect auxiliary results used in the proof of Theorem~\ref{thm_code_wt}, in the following Lemma. For~\(1 \leq w \leq n\) let~\({\mathcal C}_w\) be the set of all words in~\(\mathbb{F}_q^{n}\) with weight~\(w.\) The following result estimates the number of words of a given weight present in a linear code.
\begin{lemma}\label{mu_w_lem} We have:\\
\((a)\) For~\(1 \leq w \leq n\) let~\(N_w\) be the set of words of the random code~\({\mathcal G}\) present in~\({\mathcal C}_w.\) We have that the mean and the variance satisfy
\begin{equation}\label{mean_est}
\mu_w := \mathbb{E}N_w = {n \choose w} \cdot (q-1)^{w} \cdot \frac{q^{k}-1}{q^{n}} \text{ and }var(N_w) \leq (2q+1) \cdot \mu_w,
\end{equation}
respectively.\\
\((b)\) Let~\(\lambda_w\) and~\(\mu_w\) be as in~(\ref{mu_w_mds}) and~(\ref{mean_est}), respectively. If~\(q \geq n\) then for each~\(n-k+1 \leq w \leq n\) we have that
\begin{equation}\label{lam_mu_comp}
\left(1-\frac{1}{q}\right)\cdot \left(1-\frac{w-1}{q}\right) \leq \frac{\lambda_w}{\mu_w} \leq \frac{1}{1-\frac{w}{q}}.
\end{equation}
\end{lemma}
From~(\ref{mean_est}) in part~\((a),\) we get the intuitive result that the number of words of weight~\(w\) in a linear code is concentrated around its mean. From part~\((b)\) we see that if~\(q\) is much larger than~\(n,\) then~\(\frac{w}{q} = o(1)\) and so~\(\lambda_w\) is approximately equal to~\(\mu_w.\)

\emph{Proof of Lemma~\ref{mu_w_lem}\((a)\)}:  We first obtain the expression for~\(\mu_w.\) For any fixed non-zero vector~\(\mathbf{x} \in \mathbb{F}_q^{k},\) the random vector~\(\mathbf{x} \cdot \mathbf{G} \) is uniform in~\(\mathbf{F}_q^{n}\) and so for any vector~\(\mathbf{y} \in {\mathcal C}_w\) we have that~\(\mathbb{P}\left(\mathbf{x} \cdot \mathbf{G} = \mathbf{y}\right) = \frac{1}{q^{n}}.\) The relation for~\(\mu_w\) in~(\ref{mean_est})  then follows from the fact that the number of words of weight~\(w\) equals~\(\#{\mathcal C}_w = {n \choose w} \cdot (q-1)^{w}\)
and the fact that there are~\(q^{k}-1\) non-zero vectors in~\(\mathbf{F}_q^{k}.\) To estimate the variance of~\(N_w\) we write~\(N_w = \sum_{\mathbf{x} \in \mathbb{F}_q^{k} \setminus \{0\}} \ind\left(A(\mathbf{x})\right)\)
where~\(A(\mathbf{x})\) is the event that the vector~\(\mathbf{x} \cdot \mathbf{G}  \in {\mathcal C}_w.\)
We then get that
\begin{equation}\label{var_nw}
var(N_w) = \sum_{\mathbf{x}} \Delta(\mathbf{x}) + \sum_{\mathbf{x}_1 \neq \mathbf{x}_2} \beta(\mathbf{x}_1,\mathbf{x}_2)
\end{equation}
where~\(0 \leq \Delta(\mathbf{x}) := \mathbb{P}\left(A(\mathbf{x})\right)-\mathbb{P}^2(A(\mathbf{x})) \leq \mathbb{P}\left(A(\mathbf{x})\right)\) and
\begin{eqnarray}
|\beta(\mathbf{x}_1,\mathbf{x}_2)| &:=& \left|\mathbb{P}\left(A(\mathbf{x}_1) \bigcap A(\mathbf{x}_1)\right)-\mathbb{P}(A(\mathbf{x}_1))\mathbb{P}(A(\mathbf{x}_2)) \right|\nonumber\\
&\leq& \mathbb{P}\left(A(\mathbf{x}_1) \bigcap A(\mathbf{x}_1)\right)+\mathbb{P}(A(\mathbf{x}_1))\mathbb{P}(A(\mathbf{x}_2)) \leq 2\mathbb{P}(A(\mathbf{x}_1)). \label{beta_nw}
\end{eqnarray}

It is well-known that if~\(\mathbf{x}_1\) is not a multiple of~\(\mathbf{x}_2\) then the events~\(A(\mathbf{x}_1) \) and~\(A(\mathbf{x}_2)\) are independent (see for example, Chapter~\(7,\) Problem~\(P.7.18,\) pp. 175, (Zamir~\cite{ram}). Therefore for each~\(\mathbf{x}_1\) there are at most~\(q\) values of~\(\mathbf{x}_2\) for which~\(\beta(\mathbf{x}_1,\mathbf{x}_2) \neq 0.\)
Thus~\(var(N_w) \leq \sum_{\mathbf{x}} \mathbb{P}\left(A(\mathbf{x})\right) + 2q \sum_{\mathbf{x}_1} \mathbb{P}\left(A(\mathbf{x}_1)\right) = (2q+1) \mu_w\)
and this proves the variance estimate in~(\ref{mean_est}).~\(\qed\)

\emph{Proof of Lemma~\ref{mu_w_lem}\((b)\)}: We begin by showing that if~\(n-k+1 \leq w \leq n\) then
\begin{equation}\label{lam_w_bds}
\left(1-\frac{1}{q}\right)\cdot \left(1-\frac{w-1}{q}\right) \leq \frac{\lambda_w}{\left(\frac{{n \choose w} \cdot q^{w}}{q^{n-k}}\right)} \leq 1-\frac{1}{q}.
\end{equation}
To prove the upper bound in~(\ref{lam_w_bds}) we write~\(\lambda_w = \left(\frac{{n \choose w} \cdot q^{w}}{q^{n-k}}\right) \cdot \left(1-\frac{1}{q}\right) \cdot \theta(w)\)
where~\(\theta(w) := \sum_{j=0}^{w-D} (-1)^{j} {w-1 \choose j} \cdot \frac{1}{q^{j}}.\)
Expanding~\(\theta(w)\) and regrouping we get\\\(\theta(w) = 1- (t_1(w)+t_3(w) + t_5(w)+\ldots) \) where
\begin{equation}
t_j(w) := {w-1 \choose j} \cdot \frac{1}{q^{j}}  - {w-1 \choose j+1} \cdot \frac{1}{q^{j+1}}
\end{equation}
for all~\(j\) if~\(w-D\) is odd. If~\(w-D\) is even, then an analogous expansion holds with the distinction that the final~\(t_j(w)\) term is simply~\({w-1 \choose j} \cdot \frac{1}{q^{j}}.\) For simplicity we assume below that~\(w-D\) is odd and get~\(t_j(w) = {w-1 \choose j} \cdot \frac{1}{q^{j}}\left(1-\frac{r_j(w)}{q}\right) \) where~\(r_j(w) := \frac{{w-1 \choose j+1} }{{w-1 \choose j}} = \frac{w}{j+1}-1.\) Thus~\(|r_j(w)| \leq n\) and since~\(q \geq n\) we get that~\(t_j(w) \geq 0.\) This implies that~\(\theta(w) \leq 1\) and so we get the upper bound in~(\ref{lam_w_bds}).

For the lower bound in~(\ref{lam_w_bds}) we write~\(\theta(w)  = 1-\frac{w-1}{q} + t_2(w) + t_4(w) + \ldots \) and use~\(t_j(w) \geq 0\) to get that~\(\theta(w) \geq 1-\frac{w-1}{q}.\) This proves the lower bound in~(\ref{lam_w_bds}). Finally to prove~(\ref{lam_mu_comp}), we write~\(\mu_w =\frac{{n \choose w}q^{w}}{q^{n-k}}\cdot \left(1-\frac{1}{q}\right)^{w} \cdot \left(1-\frac{1}{q^{k}}\right)\)
and use~\(\left(1-\frac{1}{q}\right)^{w} \geq 1-\frac{w}{q}\) and~\(1-\frac{1}{q^{k}} \geq 1-\frac{1}{q}\) to get that
\begin{equation}\label{mu_bds}
\left(1-\frac{1}{q}\right) \cdot \left(1-\frac{w}{q}\right) \leq \frac{\mu_w}{\frac{{n \choose w}q^{w}}{q^{n-k}}} \leq 1.
\end{equation}
Together with~(\ref{lam_w_bds}) we then get~(\ref{lam_mu_comp}).~\(\qed\)

\subsection*{Proof of Theorem~\ref{thm_code_wt}}
We first estimate the probability of occurrence of property~\((a1).\) Recalling that~\(\mu_w = \mathbb{E}N_w\) is the expected number of words of weight~\(w\) in the random code~\({\mathcal G}\) (see Lemma~\ref{mu_w_lem}) and using Stirling's approximation we have that
\begin{equation}\label{mu_w_orig}
\mu_w \leq \frac{{n \choose w}q^{w}}{q^{n-k}} \leq 4en \cdot q^{\frac{nH\left(\frac{w}{n}\right)}{\log{q}}} \cdot \frac{q^w}{q^{n-k}} \leq 4en \cdot q^{\frac{n}{\log{q}}} \cdot \frac{q^w}{q^{n-k}}.
\end{equation}
Setting~\(w_{low} := n-k-\frac{2n}{\log{q}},\) we see for all~\(1 \leq w \leq w_{low}\) that~\(\mu_w  = \mathbb{E}N_w \leq \frac{4en}{q^{\frac{n}{\log{q}}}} = 4en \cdot e^{-n}.\) Therefore if~\(F_{low}\) is the event that the property~\((a1)\) in the statement of the Theorem holds, then we get by the union bound that
\begin{equation}\label{low_weights}
\mathbb{P}(F^c_{low}) = \mathbb{P}\left(\bigcup_{1 \leq w \leq w_{low}} \{N_w \geq 1\}\right) \leq \sum_{w=1}^{w_{low}} \mathbb{E}N_w \leq w_{low} \cdot 4en \cdot e^{-n} \leq 4en^2 \cdot e^{-n}
\end{equation}
since~\(w_{low} \leq n.\)

Next we study property~\((a2)\) for weights~\(w \geq w_{up} = n-k+5.\) First we show that
\begin{equation}\label{mu_w_est}
\mu_w \geq \frac{{n \choose w}q^{w}}{q^{n-k}}\cdot \left(1-\frac{w}{q}\right) \cdot \left(1-\frac{1}{q^{k}}\right) \geq \frac{n^5}{4}
\end{equation}
for all~\(n\) large.  the first bound in~(\ref{mu_w_est}) follows from~(\ref{mu_bds}).
For~\(w \geq n-k+5\) we have that~\(\frac{{n \choose w}q^{w}}{q^{n-k}} \geq {n \choose w} \cdot q^5 \geq n^5\)
since~\(q \geq n\) for all~\(n\) large, by~(\ref{k_q_cond}). Also from~(\ref{k_q_cond}) we see that~\(1-\frac{w}{q} \geq 1-\frac{n}{q} \geq \frac{1}{2}\) and~\(1-\frac{1}{q^{k}} \geq \frac{1}{2}\) for all~\(n\) large. This proves the final bound in~(\ref{mu_w_est}).

From Chebychev's inequality, the variance estimate in~(\ref{mean_est}) and the above estimate~(\ref{mu_w_est}), we therefore get that
\begin{equation}\label{enw2}
\mathbb{P}\left(|N_w -\mu_w| \geq \frac{\mu_w}{n}\right)  \leq \frac{n^2var(N_w)}{\mu_w^2} \leq \frac{n^2(2q+1)}{\mu_w} \leq \frac{4(2q+1)}{n^3}.
\end{equation}
If~\(\mu_w\left(1-\frac{1}{n}\right) \leq N_w\leq \mu_w\left(1+\frac{1}{n}\right),\) then using the bounds~(\ref{lam_mu_comp}), we see that
\[ \lambda_w \left(1-\frac{w}{q}\right) \cdot \left(1-\frac{1}{n}\right) \leq N_w\leq \lambda_w \frac{\left(1+\frac{1}{n}\right)}{\left(1-\frac{1}{q}\right) \cdot \left(1-\frac{w-1}{q}\right)}\] and using~(\ref{k_q_cond}) and the fact that~\(w \leq n\) we get that
\begin{equation}
N_w \leq \lambda_w \left(1+\frac{1}{n}\right) \cdot \left(1+\frac{2}{q}\right) \cdot \left(1+\frac{2n}{q}\right) \leq \lambda_w \left(1+\frac{3n}{q}\right) \nonumber
\end{equation}
for all~\(n\) large. Similarly we also get that~\(N_w \geq \lambda_w\left(1-\frac{3n}{q}\right)\) for all~\(n\) large.

Therefore if~\(F_{up}\) denotes the event that property~\((a2)\) in the statement of the Theorem holds, then from~(\ref{enw2}) and the union bound we get that~\(\mathbb{P}(F^{c}_{up})\) is bounded above by~\(\sum_{w =w_{up}}^{n} \mathbb{P}\left(|N_w-\mu_w| \geq \frac{\mu_w}{n}\right) \leq \frac{4(2q+1)}{n^2}.\) Thus~\(\mathbb{P}(F_{up}) \geq 1-\frac{4(2q+1)}{n^2}\)
and combining this with~(\ref{low_weights}) we get that
\begin{equation}\label{eq_a}
\mathbb{P}(F_{up} \cap F_{low}) \geq 1-4en^2 \cdot e^{-n} - \frac{4(2q+1)}{n^2} \geq 1-\frac{9q}{n^2}
\end{equation}
for all~\(n\) large using the fact that~\(q \geq n\) (see statement of Theorem). If~\(F_{full}\) denotes the event that the matrix~\(\mathbf{G}\) has full rank then  we show below that
\begin{equation}\label{eq_b}
\mathbb{P}(F_{full}) \geq 1-\frac{2}{q^{n-k}} \geq \frac{1}{2}
\end{equation}
since~\(q \geq 2\) and~\(n-k \geq 2\) (see statement of Theorem).

The ratio of the sets~\({\mathcal Q}\) and~\({\mathcal P}\) defined in the statement of the Theorem is therefore simply
\begin{equation}\label{q_rat}
\frac{\#{\mathcal Q}}{\#{\mathcal P}} = \frac{\mathbb{P}(F_{full} \cap F_{up} \cap F_{low})}{\mathbb{P}(F_{full})} \geq 1- \frac{\mathbb{P}(F^c_{up} \cup F^c_{low})}{\mathbb{P}(F_{full})}
\end{equation}
using~\(\mathbb{P}(A \cap B) \geq \mathbb{P}(A) - \mathbb{P}(B^c)\) with~\(A = F_{full}\) and~\(B = F_{up} \cap F_{low}.\)
Plugging~(\ref{eq_a}) and~(\ref{eq_b}) into~(\ref{q_rat}) we get that~\(\frac{\#{\mathcal Q}}{\#{\mathcal P}} \geq 1- \frac{18q}{n^2}\)
and this proves~(\ref{q_est}).

It remains to prove~(\ref{eq_b}).  Let~\(\mathbf{V}_i,1 \leq i \leq k\) be the independent and identically distributed (i.i.d.) vectors chosen uniformly randomly from~\(\mathbb{F}_q^{n}\) that form the rows of the matrix~\(\mathbf{G}.\)  For~\(1 \leq i \leq k\) let~\(E_i\) be the event that the vectors~\(\mathbf{V}_j,1 \leq j \leq i\) are linearly independent so that~\(\mathbb{P}(E_1) = 1.\)
For~\(i \geq 2,\) we note that the event~\(E_i = \bigcap_{1 \leq j \leq i} E_j\) and write
\begin{equation}\label{cond_ei}
\mathbb{P}(E_i) = \mathbb{P}\left(\bigcap_{1 \leq j \leq i} E_i\right) = \mathbb{E}\left(\ind(E_{i-1}) \cdot \mathbb{P}\left(E_i \left|\right.\mathbf{V}_j,1 \leq j \leq i-1\right)\right).
\end{equation}
If~\(E_{i-1}\) occurs, the size  of the space spanned by the vectors~\(\mathbf{V}_j, 1 \leq j \leq i-1\) is~\(q^{i-1}\) and so the event~\(E_i\) occurs if and only if we choose~\(\mathbf{V}_i\) from amongst the remaining~\(q^{n}-q^{i-1}\) vectors. Therefore from~(\ref{cond_ei}) we get that~\(\mathbb{P}(E_i) = \left(\frac{q^{n}-q^{i-1}}{q^{n}}\right)\mathbb{P}(E_{i-1})\) and
continuing iteratively, we get that
\begin{equation}\label{ed_est}
\mathbb{P}(E_k) = \prod_{j=1}^{k-1}\left(1-\frac{q^{j}}{q^{n}}\right) \geq 1-\frac{1}{q^{n}}\sum_{j=1}^{k-1}q^{j} = 1-\frac{q^{k}-1}{q^{n}(q-1)} \geq 1-\frac{2}{q^{n-k}},
\end{equation}
using the fact that~\(q \geq 2\) and so~\(\frac{q^{k}-1}{q^{n}(q-1)} \leq \frac{q^{k}}{q^{n}(q-1)} \leq \frac{2}{q^{n-k}}.\) This proves~(\ref{eq_b}).~\(\qed\)

\emph{Acknowledgement}: I thank Professors V. Guruswami, C. R. Subramanian and the referees for crucial comments that led to an improvement of the paper. I also thank IMSc for my fellowships.



\begin{thebibliography}{6}
%

\bibitem{huff}  Huffman, W. C., Pless, V. : Fundamentals of Error Correcting Codes. Cambridge University Press, (2003).

\bibitem{tol} Tolhuizen, L. : On Maximum Distance Separable Codes Over Alphabets of Arbitrary Size. Proceedings of the IEEE International Symposium on Information Theory (ISIT). pp. 926--930 (2005).

\bibitem{macw} MacWilliams, F. J., Sloane, N. J. A. : The Theory of Error Correcting Codes. vol. 16. Amsterdam, The Netherlands: North-Holland, (1977).

\bibitem{ezer} Ezerman, M. F., Grassl, M., Sole, P.  : The Weights in MDS Codes. IEEE Transactions on Information Theory, vol. 57, pp. 392--396, (2011).

\bibitem{alder} Alderson, T. L. :  On the Weights of General MDS Codes: IEEE Transactions on Information Theory, vol. 66, pp. 5414--5418, (2019).

\bibitem{ram} Zamir, R. : Lattices Coding for Signals and Networks. Cambridge University Press, (2014).



\end{thebibliography}
\end{document}